\documentclass[twocolumn,amsmath,amssymb,nofootinbib,showpacs,pra]{revtex4}

\ifx\pdfoutput\undefined
\usepackage{graphicx,color}
\else
\usepackage[pdftex]{graphicx,color}
\fi

\usepackage{amsmath}
\usepackage{amssymb}
\usepackage{amsfonts}
\usepackage{amsthm}
\usepackage{braket}
\usepackage{comment}

\newcommand{\myeqnref}[1]{Eq.~\eqref{#1}}

\newcommand{\figlength}{\linewidth}

\newtheorem{theorem}{Theorem}
\newtheorem{lemma}{Lemma}

\begin{document}

\title{Security proof of a three-state quantum key distribution protocol without rotational symmetry}

\date{\today}


\author{Chi-Hang Fred Fung}
 \email{cffung@comm.utoronto.ca}
%\affiliation{%
%Dept. of Electrical and Computer Engineering, \\
%University of Toronto, Toronto, Ontario M5S 3G4, Canada
%}%
\author{Hoi-Kwong Lo}
 \email{hklo@comm.utoronto.ca}
\affiliation{%
Center for Quantum Information and Quantum Control,\\
Dept. of Electrical and Computer Engineering and Dept. of Physics,\\
University of Toronto, Toronto, Ontario M5S 3G4, Canada
}%
%\affiliation{

\begin{abstract}
Standard security proofs of quantum key distribution (QKD) protocols often rely on symmetry arguments.
In this paper,
we prove the security of a
three-state protocol that {\it does not} possess rotational symmetry.
% which is present in many QKD protocols and thus simplifies their security proofs significantly.
%We prove the security of a quantum key distribution (QKD) protocol involving
The three-state QKD protocol we consider involves
three qubit states,
% $\{ \ket{0_z}, \ket{1_z}, \ket{0_z}+\ket{1_z} \}$,
where the first two states, $\ket{0_z}$ and $\ket{1_z}$, can contribute to key generation and the third state, $\ket{+}=(\ket{0_z}+\ket{1_z})/\sqrt{2}$, is for channel estimation.
This protocol has been proposed and implemented experimentally in some frequency-based QKD systems where the three states can be prepared easily.
%In order to a
%This protocol is mathematically equivalent to a few previously proposed protocols which have been implemented experimentally.
Thus, by founding on the security of this three-state protocol, we prove that these QKD
schemes are, in fact, unconditionally
secure against any attacks allowed by quantum mechanics.
%This protocol is mathematically equivalent to a few protocols implemented in practice, specifically, a time-bin-based scheme and a frequency-based scheme.
%By establishing
%Thus, by founding on the security of this three-state protocol, these QKD implementations can be secure against any attacks allowed by quantum mechanics.
The main task in our proof is to upper bound the phase error rate of the qubits given the bit error rates observed.
Unconditional security can then be proved not only for the ideal case of a single-photon source and perfect detectors, but also for the realistic case of a phase-randomized weak coherent light source and imperfect threshold detectors.
%
%We note that this three-state protocol does not possess rotational symmetry, which is present in many QKD protocols and simplifies their security proofs significantly.
%
Our result on the phase error rate upper bound is independent of the loss in the channel.
Also, we compare the three-state protocol with the BB84 protocol.
For the single-photon source case,
our result proves that the BB84 protocol strictly tolerates a higher quantum bit error rate than the three-state protocol;
while for the coherent-source case,
the BB84 protocol achieves a higher key generation rate and secure distance than the three-state protocol when a decoy-state method is used.
\end{abstract}

\maketitle

\section{Introduction}
Quantum key distribution (QKD) \cite{Bennett1984,Ekert1991} allows two distant parties to expand a previously shared secret key by sending quantum states through a quantum channel.
The most well-known QKD protocol is the BB84 protocol \cite{Bennett1984}, which has been proved unconditionally secure against any attacks allowed by quantum mechanics \cite{Mayers2001,Biham2000,Lo1999,Shor2000,Inamori2005,Gottesman2004,Koashi2005b,Gottesman2003,Chau2002}.
Standard security proofs of many QKD protocols, including the BB84 protocol,
the SARG04 protocol \cite{Scarani2004,Branciard2005,Tamaki2006,Fung2006},
the symmetric three-state protocol \cite{Phoenix2000,Boileau2005}, and
the generalized rotationally symmetric protocol \cite{Koashi2005a,Shirokoff2006},
often rely on rotational symmetries.
In this paper, we prove the security of a QKD protocol that {\it does not} possess rotational symmetry.
%
%In this paper,
%The protocol we consider consists of three states and prove it unconditional security rigourously.
%The protocol is mathematically equivalent to
The protocol involves Alice sending one of the three qubit states $\{ \ket{0_z}, \ket{1_z}, (\ket{0_z}+\ket{1_z})/\sqrt{2} \}$ to Bob, where the first two states
are for key generation,
%both key generation and channel estimation
 and the third state is for channel estimation.
%can contribute to key generation and the third is for channel estimation.
Note that this protocol is similar to the BB84 protocol in that they share the same three qubit states.
In fact, in practical implementations of the BB84 protocol, when one of the four laser sources is out of operation (due to, for example, malfunctioning), the QKD scheme implemented becomes the three-state protocol that we consider in this paper.
The security proof of the three-state protocol analyzed in this paper then assures that even the handicapped BB84 protocol can still be secure in these situations \cite{HarringtonMisc1}.
This three-state protocol has also been proposed and implemented in some frequency-based QKD systems \cite{Molotkov1996,Molotkov1998,Shi2000}\footnote{Note that some of these systems actually prepare the state $\ket{0_z}+\ket{1_z}$ instead, which has a different normalization (without the factor of $\sqrt{2}$) than the one we consider.  These QKD systems are not qubit-based and thus our security proof is not directly applicable to them.}.
In these frequency-based systems, %a logic $0$ ($1$)
the state $\ket{0_z}$ ($\ket{1_z}$)
is represented by a pulse in frequency $\omega_0$ ($\omega_1$), while
%a channel estimation signal
the state $(\ket{0_z}+\ket{1_z})/\sqrt{2}$
is represented by a pulse in a superposition of the two frequencies.
In these systems, it is relatively easy to generate the three states and thus the three-state protocol is well suited for these systems.
In order to understand the security of these systems, a rigourous security analysis of the three-state protocol is in order, and it is the purpose of this paper to provide such an analysis.
We note that a similar protocol has been proposed and implemented in some time-bin-based QKD systems \cite{Brendel1999,Thew2002,Tittel2001,Gisin2004,Stucki2005}.
In one time-bin-based scheme \cite{Stucki2005},
each %key-generating
signal is associated with two time positions
 and there are three different signals.
A logic $0$ ($1$) is represented by a light pulse in the first (second) position and no pulse in the other position;
 while
a decoy signal is represented by a superposition of a pulse in the first position and a pulse in the second position.
The channel is estimated by checking the coherence between two consecutive non-empty pulses appearing within or across the bit separations.
This gives rise to the difference between this protocol and the one we consider in this paper.
If only the coherence within the bit separations was checked, then it is equivalent to our protocol.
Thus, the analysis in the paper does not directly apply to this particular time-bin-based scheme.
On the other hand, the result of this paper suggests that even if only the coherence within the bit separations is checked, unconditional security can still be established, thus making it unnecessary to check for the across-the-bit coherence for the sake of achieving unconditional security.
This means that a secure time-bin-based scheme can be
built by implementing
%implemented by realizing
the three-state protocol analyzed in this paper,
where the channel-estimation state is realized by checking the within-the-bit coherence.

% On the other hand, other protocols are being proposed due to their suitability in certain implementations.
% In particular, a protocol employing three states have been proposed and implemented \cite{Gisin2004,Stucki2005,Molotkov1996,Molotkov1998,Shi2000}.
% %three-state protocol is proposed in \cite{a}.
% This protocol lends itself to easy implementation in time-bin-based QKD systems \cite{Brendel1999,Thew2002,Tittel2001} and frequency-based QKD systems \cite{Molotkov1996,Molotkov1998,Shi2000}, and thus is a natural choice for these systems.
% Specifically, the protocol is mathematically equivalent to Alice sending one of the three qubit states $\{ \ket{0_z}, \ket{1_z}, \ket{0_z}+\ket{1_z} \}$ to Bob, where the first two states can contribute to key generation and the third is for channel estimation.
% In a time-bin system, each signal is associated with two time positions and there are three different signals.
% A logic $0$ ($1$) is represented by a light pulse in the first (second) position and no pulse in the other position, while
% a channel estimation signal is represented by a superposition of a pulse in the first positions and a pulse in the second position.
% Similarly, in a frequency-based system, a logic $0$ ($1$) is represented by a pulse in frequency $\omega_0$ ($\omega_1$), while
% a channel estimation signal is represented by a pulse in a superposition of the two frequencies.
% Although both types of systems are quite different in implementation, they can be viewed as realizing the same protocol involving the three aforementioned qubit states.
In this paper, we prove the unconditional security of the three-state protocol not only for the case of a single-photon source, but also for the case of a phase-randomized weak coherent-state source.
Essentially, the reason that the protocol is secure is because the information gain by an eavesdropper implies disturbance in the signals received by a legitimate receiver.
Here, our main task is to make this argument rigourous and quantitative.
To do this, %our main task is to
we upper bound the phase error rate of the key-generating qubits using the bit error rates of the key-generating qubits and the channel-estimation qubits (c.f. \myeqnref{eqn-epsol1}) with the assumption that a single-photon source is used.
Once the phase error rate is estimated, we may establish the security of the protocol by applying Shor-Preskill's argument \cite{Shor2000} when a single-photon source is used, and
by applying the result of Gottesman-Lo-L{\"{u}}tkenhaus-Preskill (GLLP) \cite{Gottesman2004} and the decoy-state method \cite{Hwang2003,Lo2005,Ma2005b,Wang2005a,Wang2005b,Harrington2005,Zhao2006,Zhao2006b,Peng2006,Rosenberg2006} when a coherent light source is used.
We remark that our result on the phase error rate upper bound is independent of the loss in the channel, similar to the BB84 protocol.

% We note that the protocol we consider in this paper is different from the three-state protocol considered in \cite{Boileau2005}.
% Specifically, the three states in the protocol we consider have no rotational symmetry in the Bloch sphere, while those in \cite{Boileau2005} have.
% Protocols with rotational symmetry were also studied by Koashi \cite{Koashi2005a} and by us \cite{Shirokoff2006}.
% Due to the rotational symmetry in these protocols, security proofs of them can be greatly simplified.

The paper is organized as follows:
We first describe the three-state protocol in Section~\ref{sec-protocol}.
In Section~\ref{sec-phaseerrorrate}, we upper bound the phase error rate of the key-generating qubits.
This upper bound can then be used to compute the key generation rate for both the ideal case and the realistic case in Section~\ref{sec-keyrate}.
We finally conclude in Section~\ref{sec-conclusion}.

\section{The protocol \label{sec-protocol}}

In this section, we outline the three-state protocol in prepare-and-measure version which is how it can be implemented in reality without a quantum computer,
and in entanglement distillation protocol (EDP)-based version which is equivalent to the prepare-and-measure form and is used mainly for proving the security.
In the following, we assume that Alice and Bob are equipped with a perfect single-photon source and perfect detectors.
Also, only the qubits detected by Bob are considered, and thus the security proof of this protocol is loss independent.

We use the following notations: the eigenstates in the $Z$ basis are $\ket{0_z}$ and $\ket{1_z}$, whereas the eigenstates in the $X$ basis are $\ket{+} \triangleq (\ket{0_z} + \ket{1_z})/\sqrt{2}$ and $\ket{-} \triangleq (\ket{0_z} - \ket{1_z})/\sqrt{2}$.

\subsection{Prepare-and-measure version}

We outline the protocol as follows:
\begin{description}
%\item[1]
%Alice creates $4N$ random bits.
\item[1]
Alice chooses a random $8N(1+\delta)$-bit string ${\bf a}$, where $\delta>0$ is a small parameter.
For each bit $i$, if $a_i=0$, she transmits a state randomly chosen in the $\ket{0_z}$, $\ket{1_z}$ basis;
if $a_i=1$, she transmits $\ket{+}$.
\item[2]
Bob receives the $8N(1+\delta)$ qubits and using a random $8N(1+\delta)$-bit string ${\bf b}$
measures each qubit in the $Z$ basis (if $b_i=0$) or the $X$ basis (if $b_i=1$).
\item[3]
Alice announces ${\bf a}$ and Bob announces ${\bf b}$.
\item[4]
They discard any results where $a_i \neq b_i$.
With high probability, there are at least $4N$ bits left and $2N$ of them belong to each basis.
Alice decides $N$ bits in the $Z$ basis as the check bits and the remaining $N$ bits in the $Z$ basis as the data bits.
\item[5]
Alice and Bob announce the values of the $N$ check bits in the $Z$ basis and the $2N$ check bits in the $X$ basis.
They compute the quantum bit error rates for the two sets separately.
We denote the two quantum bit error rate (QBER) values by $e_b$ and $\alpha$, respectively.
\item[6]
They choose an error correcting code capable of correcting errors at a bit error rate of $e_b$.
Alice computes the bit error syndrome of her data bits using this code and transmits the syndrome to Bob.
Bob corrects the errors in his data bits.
\item[7]
They estimate the phase error rate $e_p$ of the data bits from $e_b$ and $\alpha$ and choose a binary block code capable of correcting errors at a rate of $e_p$.
They apply the generator matrix of the code to their data bits, producing the final secret key.
\end{description}

We remark that the data bits consist of only the key-generating qubits 
$\{\ket{0_z},\ket{1_z}\}$, while the check bits consist of all qubits, 
$\{\ket{0_z},\ket{1_z},\ket{+}\}$, of which the first two are also used for the key generation and the third is only for channel estimation.
The task is to estimate the phase error rate, $e_p$, of the data bits from the bit error rates, $e_b$ and $\alpha$, of the check bits.
Also note that this three-state protocol is very similar to the BB84 protocol.
The only difference is that the $\ket{-}$ state in BB84 is not used in this protocol.

%Alice randomly chooses whether to send a data bit or a test bit.

%data bit from $\{\ket{0},\ket{1}\}$.

%test bit is $\ket{+}$

\subsection{EDP-based version}

Now we describe the equivalent EDP-based QKD protocol.
During the quantum state transmission phase, Alice sends Bob $4N$ quantum signals through a channel controlled by an eavesdropper Eve.
Specifically, for the $l^{\text{th}}$ signal,
Alice prepares the state
\begin{eqnarray}
\ket{\Psi}_{AB}&=&\frac{1}{\sqrt{2}} \ket{0}_K \Big[ \frac{1}{\sqrt{2}} \ket{0_z}_A \ket{0_z}_B + \frac{1}{\sqrt{2}} \ket{1_z}_A \ket{1_z}_B \Big] \nonumber \\
&&+ \frac{1}{\sqrt{2}} \ket{1}_K \ket{+}_A \ket{+}_B
\end{eqnarray}
and sends system $B$ to Bob through Eve while keeping system $A$ to herself.
In the most general attack by Eve, she interacts the $4N$ signals sent by Alice and some ancilla with an unitary operation.
An output qubit from the unitary operation is then sent to Bob for the $l^{\text{th}}$ transmission.   % (see Fig.~\ref{fig-generalattack}).
% \begin{figure}
% \centering
% %\includegraphics[width=\figlength]{generalattack}
% \caption{\label{fig-generalattack}
% We consider the most general attack by Eve, where she interacts the $N$ signals sent by Alice and some ancilla with a unitary operation.
% %An output qubit of the unitary operation is then sent to Bob.
% Eve may perform some measurements at later times, possibly after the announcements of the basis information.
% }
% \end{figure}
%Bob sequentially measures his received qubits from $l=1$ to $l=N$,
We assume that Bob always uses the same basis as Alice for each qubit pair,
%using the same basis as Alice.
since the qubit pairs where Alice and Bob
measure with different bases are discarded.
%Thus, we can safely
%assume that Bob and Alice always use the same bases for each qubit
%pair.
Specifically, for the $2N$ check qubit pairs in the $\ket{1}_K$ part, Bob measures in the
$\{\ket{+},\ket{-}\}$ basis, and
since Alice always sends the $\ket{+}$ state to Bob,
he declares an error (no error) if the
measurement outcome is $\ket{-}$ ($\ket{+}$). This allows him to
compute the QBER for this part, which we denote by $\alpha$.
For the $2N$ qubit pairs in the $\ket{0}_K$ part, Alice and Bob
randomly choose $N$ of them as check qubit pairs
%sacrifice $N$ of them chosen randomly
and compare their values publicly. They both perform
$Z$ basis measurements on them and announce their measurement outcomes in order to compute the
QBER for these $N$ qubit pairs, which we denote by $e_b$.
%Because of the random
%sampling theorem, the QBER of the remaining data qubit pairs is very
%close to $e_b$ with high probability.
%Thus,
An error correcting code
capable of correcting errors up to a bit error rate of $e_b$ can be
used by Alice and Bob to remove errors in the remaining $N$ data qubit pairs, which are
then privacy amplified to produce the final key. Since the amount of
privacy amplification needed to eliminate Eve's information on the
final key is indicated by the phase error rate of the data qubit
pairs (denoted by $e_p$), Alice and Bob need to upper bound this
quantity from what they observed, $e_b$ and $\alpha$. In what
follows, we solve this problem of upper bounding $e_p$ given fixed
values of $e_b$ and $\alpha$. Once $e_p$ is obtained, the key
generation rate can easily be computed using $e_p$ and $e_b$.

\section{Upper bounding the phase error rate\label{sec-phaseerrorrate}}

In this section, we solve the main problem of upper bounding the phase error rate in the data qubit pairs, using
the bit error rates observed in the check qubit pairs.
The values $e_b$ and $\alpha$ are actually observed in the check qubit pairs, not in the data qubit pairs.
On the other hand, we are interested in the bit error rates of the data qubit pairs, not the check qubit pairs.
In order to relate $e_b$ and $\alpha$ to the data qubit pairs, we apply a random sampling argument to
%Nevertheless, using the random sampling argument, it can be argued
infer
that what is observed in the check qubit pairs is very close to what could be observed in the data qubit pairs.
% if the same measurements were performed.
Specifically, the random sampling argument can be stated as follows:
\begin{lemma}[Random sampling test (see, for example, \cite{Nielsen2000})]
Given $2N$ bits,
they are randomly divided into two sets, each containing $N$ bits.
%randomly test $N$ and obtain $c_t$ errors.
%and test the remaining $N$ bits obtaining $c_u$ errors.
%$\delta n$ errors.
Then,
\begin{eqnarray}
\text{Pr}\{c_1 < \delta N \text{ and } c_2 > (\delta + \epsilon)N\}
< \exp[-O(\epsilon^2 N)] ,
\end{eqnarray}
where $c_1$ and $c_2$ are the number of ones in the two sets, $\delta \leq 1$ is some fraction representing the number of ones, and $\epsilon >0$ is a small parameter.
\end{lemma}
Therefore, with high probability, the bit error rates of $e_b$ and $\alpha$ could be observed in the data qubit pairs.
Note that the use of classical probability argument is valid here, since the events contributing to $e_b$, $e_p$, and $\alpha$ are outcomes of a projection measurement projecting onto the states $\{\ket{0}_K \ket{\phi_{ij}}$ ,$\ket{1}_K \ket{\phi_{i'j'}}$: $i,j,i',j'=0,1\}$.
Here, $\ket{\phi_{ij}}$ are the Bell states:
\begin{eqnarray}
\label{eqn-BellStates}
\ket{\phi_{ij}} &=& \frac{1}{\sqrt2}(\ket{0i}_z+(-1)^j \ket{1\bar{i}}_z) .
\end{eqnarray}
% the Bell basis:
% $$
% \begin{aligned}
% |\phi_{00}\rangle&=\frac{1}{\sqrt2}(|00\rangle+|11\rangle)\\
% |\phi_{10}\rangle&=\frac{1}{\sqrt2}(|01\rangle+|10\rangle)\\
% |\phi_{11}\rangle&=\frac{1}{\sqrt2}(|01\rangle-|10\rangle)\\
% |\phi_{01}\rangle&=\frac{1}{\sqrt2}(|00\rangle-|11\rangle).\\
% \end{aligned}
% $$
In what follows, because of this random sampling argument, we assume that the QBERs $e_b$ and $\alpha$ are also observed in the $N$ data qubit pairs.
Now the model becomes Alice sending $N$ data qubits to Bob through Eve who may perform on them any joint operation that are consistent with $e_b$ and $\alpha$.
Since we only consider the data qubit pairs, we index them from $l=1$ to $l=N$ for simplicity.
Eve's operation on the $l^\text{th}$ data qubit pair can conveniently be represented
in the Kraus (or operator sum) form,
%by a superoperator ({\bf --check--}),
$\mathcal{E}^{(l)}(\rho)=\sum_f E^{(l,f)} \rho {E^{(l,f)}}^\dagger$,
where the set of operator $\{E^{(l,f)}:\forall f\}$ defines the mapping for the $l^\text{th}$ data qubit pair,
and all the other data qubit pairs have been traced over.
Recall that the our main problem is to upper bound $e_p$ over all Eve's operations $\mathcal{E}^{(l)}(\cdot)$ that are consistent with the observed values of $e_b$ and $\alpha$.
Essentially, there are two constraints in our optimization problem -- one associated with $e_b$ and the other with $\alpha$.
We first consider the constraint with $\alpha$ by computing the correct/incorrect probabilities associated with each data qubit pair if a measurement in the $\ket{1}_K$ basis were to be performed.
%
%Note that Eve is ignorant of which qubits are the check qubits and which are the data qubits, and
%thus, her operation is independent of this fact.
%%
% For
%the check qubit pairs $l$, Eve's operation
%
%The optimization variable for our main problem of upper bounding $e_p$ is Eve's operation.
%Recall that the our main problem is to upper bound $e_p$ over all Eve's operations that are consistent with the observed values of $e_b$ and $\alpha$.
%We want to find the operation that gives rise to the highest value of $e_p$ among the operations that are consistent with the observed values of $e_b$ and $\alpha$.
%Essentially, there are two constraints in our optimization problem -- one associated with $e_b$ and the other with $\alpha$.
%To proceed, Eve's operation on the $l^\text{th}$ data qubit pair can conveniently be represented
%by a superoperator ({\bf --check--}),
%$\mathcal{E}^{(l)}(\rho)=\sum_f E^{(l,f)} \rho {E^{(l,f)}}^\dagger$,
%where all the other data qubit pairs have been traced over.
%Now that we have Eve's operation defined, we can compute the various correct/incorrect probabilities associated with the qubits.
%check qubits and the data qubits.
%For each data qubit pair,
In this basis,
there are only two outcomes -- either that Alice sends $\ket{+}$ and Bob receives $\ket{+}$ (no bit error)
or that Alice sends $\ket{+}$ and Bob receives $\ket{-}$ (a bit error).
%a bit error occurs or no bit error occurs.
The corresponding probabilities are
%$\mathcal{E}^{(l)}()$
\begin{eqnarray}
\label{eqn-check-bit-errorprob}
p_{+-}^{(l)}&\triangleq&\text{Pr}\{\text{error at position $l$}\} \\
&=& _{KAB} \langle 1+-| \mathcal{E}^{(l)}(\ket{\Psi}\bra{\Psi}) |1+-\rangle _{KAB} \nonumber \\
\label{eqn-check-bit-noerrorprob}
p_{++}^{(l)}&\triangleq&\text{Pr}\{\text{no error at position $l$}\} \\
&=& _{KAB} \bra{1++} \mathcal{E}^{(l)}(\ket{\Psi}\bra{\Psi}) |1++\rangle _{KAB} . \nonumber
\end{eqnarray}
(The notation used is that $+-$ ($++$) in the subscript means that Alice sends $\ket{+}$ and Bob measures $\ket{-}$ ($\ket{+}$).)
To construct the first constraint, we need to relate these two probabilities for the data bits to $\alpha$.
Note that they are not related in a straightforward manner,
%there are two complications:
%(i) $\alpha$ is the {\it observed} bit error rate in the check bits while
since $\alpha$ is the observed bit error rate in the data qubit pairs (inferred from that of the check qubit pairs using the random sampling argument) while we only have probabilities of each data qubit pair on hand.
%, and also Eve may operate jointly on all qubit pairs introducing correlations into the probabilities.
%First, we use the random sampling argument to infer that
%even though
%$\alpha$ is the observed bit error rate in the check bits,
%almost the same QBER could be obtained in the data bits had the same measurements using the $k=1$ basis been performed.
%
%Now that $\alpha$ is the bit error rate in the $\ket{+}$, $\ket{-}$ basis for the data bits,
%we want to relate this {\it observed} value to the probabilities of
%
%
In this situation, we utilize Azuma's inequality \cite{Azuma1967} to establish the relation, as used similarly in \cite{Boileau2005,Tamaki2006,Fung2006}.
%To begin, we assume that Bob measures his qubits in the sequence of increasing values of $l$.
%Also, Eve's operation on the $l^\text{th}$ qubit pairs $\mathcal{E}^{(l)}(\cdot)$ is obtained by tracing over the ${l'}^\text{th}$ qubit pairs where $l'<l$ conditional on the measurement outcomes of and over the ${l''}^\text{th}$ qubit pairs where $l''>l$ unconditionally.
To proceed, we obtain Eve's operation on the $l^\text{th}$ qubit pair, $\mathcal{E}^{(l)}(\cdot)$, by tracing over the previously measured qubit pairs conditional on their measurement outcomes and unconditionally tracing over the qubit pairs to be measured later.
This means that the two probabilities in Eqs.~\eqref{eqn-check-bit-errorprob} and \eqref{eqn-check-bit-noerrorprob} are now {\it conditional} probabilities, conditional on the measurement outcomes of the previously measured qubits.
%Taking each probability
Considering each event separately, Azuma's inequality asserts that the sum of the error (no error) probabilities over all qubits is close to the observed counts of the error (no error) events.
Mathematically, it means that
\begin{eqnarray}
\text{Pr} \left[
\left|
\frac
{c_{+\pm} - \sum_{l=1}^{N} p_{+\pm}^{(l)}}
{N}
\right|
\geq \epsilon
\right]
\leq 2 \exp^{-N \epsilon^2/2},
\end{eqnarray}
where $c_{+-}$ ($c_{++}$) is the observed counts of the error (no error) events, $p_{+-}^{(l)}$ ($p_{++}^{(l)}$) is error (no error) probability for the $l^\text{th}$ qubit pairs given by Eq.~\eqref{eqn-check-bit-errorprob} (Eq.~\eqref{eqn-check-bit-noerrorprob}),
and $\epsilon>0$ is a small quantity.
Note that this probability drops exponentially as $N$ increases.
Now, since $\alpha=c_{+-}/(c_{+-}+c_{++})$ by definition, it is easy to relate $p_{+\pm}^{(l)}$ to $\alpha$ as $N$ goes to infinity as follows:
\begin{eqnarray}
\label{eqn-alpha-azuma1}
\alpha &=&
\frac
{\sum_{l=1}^{N} p_{+-}^{(l)}}
{\sum_{l=1}^{N} p_{+-}^{(l)}+p_{++}^{(l)}}.
\end{eqnarray}
Note that no actual measurement in the $\{\ket{+},\ket{-}\}$ basis is performed on the data qubit pairs (only measurements in the $\{\ket{0_z},\ket{1_z}\}$ basis are performed on them) and thus we have no measurement outcomes of the earlier qubit pairs to explicitly form $\mathcal{E}^{(l)}(\cdot)$.
%However, all we need to show is that
Nevertheless, Eq.~\eqref{eqn-alpha-azuma1} holds for any measurement outcomes, and
there is no need to know what these outcomes are.
Note that as an alternative to Azuma's inequality, the quantum de Finetti theorems \cite{Renner2005b,Koenig2005,Christandl2006,DCruz2006} may be used to argue that the entanglement between a subset of the randomly permuted qubit pairs vanishes, establishing \myeqnref{eqn-alpha-azuma1} also.
In this case, a sublinear number of qubit pairs have to be discarded.

By the same token, the second constraint of our optimization problem associated with $e_b$ can be constructed in a similar way.
In this case,
there are four possibilities associated with the data qubit pairs:
no error, a bit error, a phase error, and both types of errors.
Thus, they give rise to the following four probabilities:
\begin{eqnarray}
%\label{eqn-check-bit-errorprob}
q_{rs}^{(l)} &\triangleq& _K \langle 0| _{AB} \bra{\phi_{rs}} \mathcal{E}^{(l)}(\ket{\Psi}\bra{\Psi}) |\phi_{rs}\rangle _{AB} |0\rangle _K, \\
&& \hspace{1cm} r,s=0,1, \nonumber
%\label{eqn-check-bit-noerrorprob}
%p_{++}^{(l)}\triangleq\text{Pr}\{\text{no error in check bit at position $l$}\} &=& \bra{+} \mathcal{E}^{(l)}(\ket{\Psi_\text{c}}\bra{\Psi_\text{c}}) \ket{+} .
\end{eqnarray}
where $\ket{\phi_{rs}}$ are the Bell states defined in \myeqnref{eqn-BellStates}.
Applying Azuma's inequality gives
\begin{eqnarray}
\label{eqn-eb-azuma1}
e_b &=&
\frac
{\sum_{l=1}^{N} q_{10}^{(l)}+q_{11}^{(l)}}
{\sum_{l=1}^{N} q_{00}^{(l)}+q_{10}^{(l)}+q_{11}^{(l)}+q_{01}^{(l)}} \\
\label{eqn-ep-azuma1}
e_p &=&
\frac
{\sum_{l=1}^{N} q_{01}^{(l)}+q_{11}^{(l)}}
{\sum_{l=1}^{N} q_{00}^{(l)}+q_{10}^{(l)}+q_{11}^{(l)}+q_{01}^{(l)}} .
\end{eqnarray}
Therefore, our optimization problem becomes maximizing $e_p$ given in \myeqnref{eqn-ep-azuma1} over Eve's operations $\mathcal{E}^{(l)}(\cdot)$ subject to Eqs.~\eqref{eqn-alpha-azuma1} and \eqref{eqn-eb-azuma1}.
To simplify the problem, by using the parameterization $E^{(l,f)}=a_I^{(l,f)} I + a_X^{(l,f)} X + a_Y^{(l,f)} Y + a_Z^{(l,f)} Z$ and
explicitly evaluating $p_{+\pm}^{(l)}$ and $q_{rs}^{(l)}$, we re-write the maximization problem as follows:
\begin{eqnarray}
\label{eqn-prob1}
&&\hspace{-1.1cm}\text{maximize} \nonumber \\
%\max &
\label{eqn-ep1}
e_p &=& \frac{\displaystyle\sum_{l,f} \left|a_Z^{(l,f)}\right|^2 + \left|a_Y^{(l,f)}\right|^2}{\displaystyle\sum_{l,f} \left|a_I^{(l,f)}\right|^2 + \left|a_X^{(l,f)}\right|^2 + \left|a_Y^{(l,f)}\right|^2 + \left|a_Z^{(l,f)}\right|^2} \;
\\
%s.t.&
&&\hspace{-1.1cm}\text{subject to} \nonumber \\
\label{eqn-eb1}
e_b &=& \frac{\displaystyle\sum_{l,f} \left|a_X^{(l,f)}\right|^2 + \left|a_Y^{(l,f)}\right|^2}{\displaystyle\sum_{l,f} \left|a_I^{(l,f)}\right|^2 + \left|a_X^{(l,f)}\right|^2 + \left|a_Y^{(l,f)}\right|^2 + \left|a_Z^{(l,f)}\right|^2}
\\
%&
\label{eqn-alpha1}
\alpha &=& \frac{\displaystyle\sum_{l,f} \left|i a_Y^{(l,f)} - a_Z^{(l,f)}\right|^2}{\displaystyle\sum_{l,f} \left|i a_Y^{(l,f)} - a_Z^{(l,f)}\right|^2 + \left|a_I^{(l,f)} + a_X^{(l,f)}\right|^2}
%\myeqnref{eqn-eb1}
%\text{ and }
%\myeqnref{eqn-alpha1}
\end{eqnarray}
where the maximization is over all $a_\beta^{(l,f)}$, $\beta=I,X,Y,Z$.
Note that the summation over all the qubit pairs $l$ in this problem signifies that Eve's attack is a joint attack.
%Also, the
However, the following theorem says that a collective attack by Eve is as powerful as a joint attack in the sense of causing the same bit and phase error rates $\{e_b, \alpha, e_p\}$.
Furthermore, Eve's collective attack only needs to consist of one Kraus operator.
This theorem essentially eliminates the need to consider joint attacks in upper bounding the phase error rate.
\begin{theorem}[Reduction from a joint attack to a collective attack]
For the three-state protocol, any values of the bit and phase error rates $\{e_b, \alpha, e_p\}$ achievable by any joint attack consisting of any number of Kraus operators are also achievable by a collective attack consisting of only one Kraus operator.
\end{theorem}
\begin{proof}
The idea is that any two sets
$\{a_I^{(l,f)},a_X^{(l,f)},a_Y^{(l,f)},a_Z^{(l,f)}\}$ and $\{a_I^{(l',f')},$ $a_X^{(l',f')},$ $a_Y^{(l',f')},$ $a_Z^{(l',f')}\}$
can be combined into one set without changing the values of $e_b$, $\alpha$, and $e_p$ (see Appendix~\ref{app-onlyonef}).
%$\{a_I^{(l,f)},a_X^{(l,f)},a_Y^{(l,f)},a_Z^{(l,f)}\}$ and $\{a_I^{(l',f')},a_X^{(l',f')},a_Y^{(l',f')},a_Z^{(l',f')}\}$
%$\{a_\beta^{(l,f)}\}$
Repeated applications of this idea can reduce any number of sets into one set.
This means that whatever values of $\{e_b,\alpha,e_p\}$ achievable by any number of sets are also achievable by just one set.
\end{proof}
%To simplify the problem further, we show
The consequence of this theorem is
that it is sufficient to consider $(l,f)$ taking on only one value (i.e. dropping the summations over $l$ and $f$)
%Thus, we may consider using just one set
in the maximization problem in Eq.~\eqref{eqn-prob1} without loss of generality.
This is an important consequence since the original maximization problem in Eq.~\eqref{eqn-prob1} involves infinitely many optimization variables ($a_\beta^{(l,f)}, \forall l,f$), and
the new maximization problem involves only four optimization variables ($a_I, a_X, a_Y, a_Z$).
This is a significant simplification in the problem.
%Note that similar reduction results were also obtained in an information-theoretic security proof \cite{Renner2005} and can also be deduced from the quantum de Finetti theorems \cite{Koenig2005,Christandl2006,DCruz2006}.
Note that the reduction from joint attacks to collective attacks was first discussed in Ref.~\cite{Lo2001}.
The idea was also implicitly used in Ref.~\cite{Gottesman2003}.
%Note that
Similar reduction results with explicit proofs
were also obtained in an information-theoretic security proof \cite{Renner2005} and can also be deduced from the quantum de Finetti theorems \cite{Renner2005b,Koenig2005,Christandl2006,DCruz2006}.
These two techniques are different from ours.
In particular,
%However,
the difference between the techniques involving the quantum de Finetti theorems and ours is that
the former requires discarding a sublinear number of qubits and ours does not require any discarding.
This difference may have practical implications when the number of qubits is finite.
Even though the number of discarded qubits in the de Finetti approximation is insignificant in the asymptotic case,
it may be significant in the finite situation.
%be relevant to practical situations where the number of signals is finite.
The difference between the information-theoretic security proof and ours is that
in the former, a collective attack is equal to a joint attack in the sense that the smooth R\'{e}nyi entropies of the states in the two attacks are roughly equal,
and in our proof, the two attacks are equal in the sense that they both cause exactly the same bit and phase error rates.
Also, we further show that it is sufficient to consider a collective attack consisting of only one Kraus operator as opposed to infinitely many Kraus operators.

\subsection{Exact upper bound}

In order to simplify the maximization problem in Eq.~\eqref{eqn-prob1}, we first write it as
\begin{eqnarray}
\max && \left( \left|a_Z\right|^2 + \left|a_Y\right|^2 \right) e_b\\
s.t.&& \left|a_X\right|^2 + \left|a_Y\right|^2 =1 \label{eqn-scalingconstraint1}\\
&& \frac{1-e_b}{e_b} = \left|a_I\right|^2 + \left|a_Z\right|^2 \\
&& \frac{1-\alpha}{\alpha} = \frac{\left|a_I+a_X\right|^2}{\left|i a_Y-a_Z\right|^2}
\end{eqnarray}
where the first constraint is introduced to fix the scaling of $a_\beta$'s, the second and third constraints are rearrangements of \myeqnref{eqn-eb1} and \myeqnref{eqn-alpha1}.
To simplify the problem further, we note that in order to maximize the objective, % $|a_Z|^2+|a_Y|^2$,
the third constraint should be taken so that $a_I$ and $a_X$ are in phase with each other and $i a_Y$ and $a_Z$ are in phase with each other.
This results in the following problem:
\begin{eqnarray}
\max && \left( \left|a_Z\right|^2 + \left|a_Y\right|^2 \right) e_b \label{eqn-prob3-obj}\\
s.t.&& \left|a_X\right|^2 + \left|a_Y\right|^2 =1 \label{eqn-prob3-con1}\\
&& \hat{e_b} := \frac{1-e_b}{e_b} = \left|a_I\right|^2 + \left|a_Z\right|^2 \label{eqn-prob3-con2}\\
&& \hat{\alpha} := \frac{1-\alpha}{\alpha} = \frac{\left|a_I\right|^2+\left|a_X\right|^2+2 \left|a_I\right| \left|a_X\right|}{\left|a_Y\right|^2+\left|a_Z\right|^2-2 \left|a_Y\right| \left|a_Z\right|} \label{eqn-prob3-con3}
\end{eqnarray}
Since the feasible region in $(|a_I|,|a_X|,|a_Y|,|a_Z|)$ is described by three constraints,
we can eliminate two of them, namely $a_I$ and $a_X$, to get one single constraint
%The three constraints can be combined into one equation
describing the feasible region in terms of $(|a_Y|,|a_Z|)$
by substituting
Eqs.~\eqref{eqn-prob3-con1} and \eqref{eqn-prob3-con2} into Eq.~\eqref{eqn-prob3-con3}:
%as follows:
\begin{eqnarray}
%e - z^2 + 1 - y^2 + 2 Sqrt[1 - y^2]Sqrt[e - z^2] == f(y^2 + z^2 - 2y z)
&(\hat{e_b} - |a_Z|^2) + (1 - |a_Y|^2) + 2 \sqrt{1-|a_Y|^2} \sqrt{\hat{e_b}-|a_Z|^2} = \nonumber \\
&\hat{\alpha} (|a_Y|^2 + |a_Z|^2 - 2|a_Y||a_Z|) .
\end{eqnarray}
%Since this is a quartic equation,
Squaring both sides gives a quartic equation, which admits four solutions for $|a_Z|$ in terms of $|a_Y|$.
However, there are only two valid solutions %for $|a_Z|$ in terms of $|a_Y|$
in the region $e_b \leq 1/2$ and $\alpha \leq 1/2$:
\begin{eqnarray}
\label{eqn-zsol}
&&|a_Z|=
\frac{1}{1+\hat{\alpha}} %\times \\
%&&
\bigg[
\hat{\alpha} |a_Y| \pm \sqrt{\hat{\alpha}(1-|a_Y|^2)} \pm \\
&&
\sqrt{-1+\hat{e_b}(1+\hat{\alpha})-|a_Y|^2(\hat{\alpha}-1) \pm 2 |a_Y| \sqrt{\hat{\alpha}(1-|a_Y|^2)}}
\bigg] , \nonumber \\
&& \hspace{.5cm}e_b, \alpha \leq 1/2,
\nonumber
\end{eqnarray}
%where the signs are $(--+), (-++), (+--), (++-)$.
where the signs are $(--+)$ and $(++-)$.
Since $|a_Z|$ is part of the objective function of the problem in \myeqnref{eqn-prob3-obj},
we want to use of the solution of $|a_Z|$ that is the largest.
Therefore, we use the solution of $|a_Z|$ with signs $(++-)$
% \begin{lemma}
% The solution of $|a_Z|$ with signs $(++-)$ is the largest.
% \end{lemma}
% \begin{proof}
% Obviously, the fourth solution $(++-)$ is larger than the third $(+--)$, and the second $(-++)$ is larger than the first $(--+)$.
% Thus, all we need to show is that the fourth is larger than the second.
% Defining $c:=2\sqrt{\hat{\alpha}(1-|a_Y|^2)}$ and $a:=-1+\hat{e_b}(1+\hat{\alpha})+|a_Y|^2(1-\hat{\alpha})$,
% this amounts to showing
% \begin{eqnarray}
% &&c/2+\sqrt{a-|a_Y|c}>-c/2+\sqrt{a+ |a_Y|c} \\
% \Leftrightarrow &&
% c^2 + 2 c \sqrt{a-|a_Y|c}-2|a_Y|c > 0 \\
% \Leftarrow &&
% c(c-2|a_Y|) > 0 \\
% \Leftrightarrow &&
% |a_Y|^2 < \frac{\hat{\alpha}}{\hat{\alpha}-1}.
% \end{eqnarray}
% The last inequality is true since we have $|a_Y|^2 \leq 1$ from \myeqnref{eqn-scalingconstraint1}.
% Thus, the $(++-)$ solution is the largest.
% \end{proof}
%Now, the problem becomes %maximizing $|a_Y|^2 + |a_Z|^2$ with $|a_Z|$ substituted from \myeqnref{eqn-zsol} with signs $(++-)$ over $|a_Y|\leq 1$.
and the problem becomes
\begin{eqnarray}
\label{eqn-probsimplified1}
\max_{|a_Y|\leq 1} && \left( \left|a_Z\right|^2 + \left|a_Y\right|^2 \right) e_b
\end{eqnarray}
where $|a_Z|$ substituted from \myeqnref{eqn-zsol} with signs $(++-)$.
This problem can be solved numerically for some fixed $e_b$ and $\alpha$ to obtain an upper bound on $e_p$ (which is the objective value of the problem).
Note that Eve can always construct an attack with $e_p=1/2$ that is as powerful as any arbitrary attack with an arbitrary $e_p \leq 1$ \cite{BoileauMisc1}.
%{\bf (cannot find reference)}.
She can construct this new attack by launching half of the time the arbitrary attack and the other half of the time the arbitrary attack with a phase flip operation.
In this way, the phase error rate of this new attack is $1/2$.
%the average of that of the original attack and
%Thus, when the phase error rate obtained by solving the problem in \myeqnref{eqn-probsimplified1} is greater than $1/2$,
%we may consider there exists some equally powerful attack with $e_p=1/2$ and
%solution to
%Note that when the resultant $e_p$ is greater than $1/2$,
%this means that the worst-case phase error rate of $1/2$ is achievable and thus one should set $e_p$ to $1/2$ in this case.

\subsection{Limiting cases\label{subsec-limitingcases}}
%{Special cases}
We need to deal with the cases that $e_b=0$, $\alpha=0$, or both separately.
For the case that $e_b=0$ and $\alpha>0$, we see from \myeqnref{eqn-eb1} that $a_X=a_Y=0$ and thus $e_p=\alpha$.
For the case that $e_b>0$ and $\alpha=0$, we see from \myeqnref{eqn-alpha1} that $a_Z=i a_Y$ and thus $e_p\leq 2 e_b$.
For the case that $e_b=0$ and $\alpha=0$, we have $a_X=a_Y=a_Z=0$ and thus $e_p=0$.
Note that the last case is consistent with the idea that information gain implies disturbance.
Since there is no disturbance in that case, no information is gained by Eve and thus $e_p=0$.

\subsection{Closed-form approximate upper bound}
It may be difficult to solve the problem in \myeqnref{eqn-probsimplified1} analytically.
Thus, in order to obtain an analytical upper bound on $e_p$, instead of using the exact value for $|a_Z|$ from \myeqnref{eqn-zsol}, we use an upper bound of $|a_Z|$ which is given by
\begin{eqnarray}
|a_Z|&\leq&
\frac{1}{1+\hat{\alpha}}
\biggr[
\hat{\alpha} |a_Y| + \sqrt{\hat{\alpha}(1-|a_Y|^2)} + \nonumber \\
&&\sqrt{-1+\hat{e_b}(1+\hat{\alpha})}
\biggr] .
\end{eqnarray}
We use this upper bound for $|a_Z|$ in the problem
%problem \myeqnref{eqn-probsimplified1},
$\max_{|a_Y|\leq 1} \left( \left|a_Z\right|^2 + 1 \right) e_b$.
Since the objective value of this problem is larger than or equal to the objective value of the original problem in \myeqnref{eqn-probsimplified1},
the solution of this problem is definitely an upper bound (but may not be tight) on $e_p$.
The solution to the approximate upper bound is
\begin{eqnarray}
e_p &\leq&
\left[
1 +
\left(
\sqrt{\frac{\hat{\alpha}}{1+\hat{\alpha}}}
+ \frac{1}{1+\hat{\alpha}} \sqrt{-1+\hat{e_b}(1+\hat{\alpha})}
\right)^2
\right] e_b \nonumber \\
&=&
\left[
1 +
\left(
\sqrt{1-\alpha}
+ \alpha \sqrt{-1+\frac{1-e_b}{e_b \alpha})}
\right)^2
\right] e_b \nonumber \\
%,  e_b,\alpha>0 \nonumber \\
&=&
%(1+(1-\alpha)^2)
\label{eqn-epsol1}
\alpha +  e_b (2-2\alpha-\alpha^2)+ \\
&&2 \sqrt{\alpha (1-\alpha) e_b (1-e_b-e_b \alpha)} %\nonumber \\
 , \: \: e_b, \alpha \leq 1/2. \nonumber
\end{eqnarray}
Note that the three special cases in Section~\ref{subsec-limitingcases} can be obtained by taking the corresponding limit in \myeqnref{eqn-epsol1}.
\begin{figure}
\centering
\includegraphics[width=\figlength]{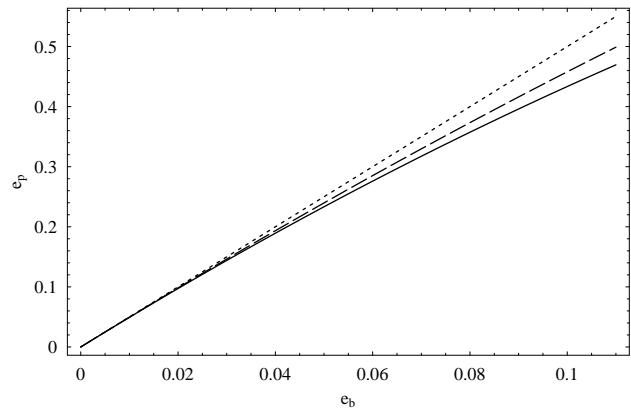}
\caption{\label{fig-compare1}
Comparison of the approximate upper bounds on $e_p$ in \myeqnref{eqn-epsol1} and \myeqnref{eqn-ep5eb} with the optimal one obtained by solving the problem in \myeqnref{eqn-probsimplified1}, assuming $e_b=\alpha$.
The solid, dashed, and dotted curves (from bottom to top) correspond to the optimal bound (\myeqnref{eqn-probsimplified1}), the general approximate bound (\myeqnref{eqn-epsol1}), and the specific approximate bound (\myeqnref{eqn-ep5eb}).
}
\end{figure}
In order to illustrate how good the approximate upper bound in \myeqnref{eqn-epsol1} is compared to the optimal one obtained by solving the problem in \myeqnref{eqn-probsimplified1} numerically,
we plot in Fig.~\ref{fig-compare1} the two bounds on $e_p$ over different values of $e_b$ assuming $e_b=\alpha$.
It can be seen that the approximate bound is very close to the optimal one, especially for small $e_b$.
%Note that when both $e_b$ and $\alpha$ are small,
Note that one may obtain another simple bound on $e_p$ from \myeqnref{eqn-epsol1} as
\begin{equation}
\label{eqn-epsimplebound1}
e_p \leq \alpha + 2e_b +2 \sqrt{e_b \alpha}.
\end{equation}
This bound is close to the bound in \myeqnref{eqn-epsol1} when both $e_b$ and $\alpha$ are small.

\subsection{Special case: $e_b=\alpha$}
For the special case $e_b=\alpha$, a linear relation between $e_b$ and the approximate $e_p$ can be derived easily.
Substituting $\alpha=e_b$ in
\myeqnref{eqn-epsimplebound1}, we get
%\myeqnref{eqn-epsol1}, we get
\begin{eqnarray}
% e_p &\leq&
% e_b \left[ 3- e_b (2+e_b) + 2 \sqrt{1-2e_b+e_b^3} \right] \\
% &\leq&
% e_b \left[ 3- e_b (2+e_b) + 2 \sqrt{1-2e_b+e_b^2} \right] \\
% &=&
% e_b (5-4e_b-e_b^2) \\
\label{eqn-ep5eb}
e_p &\leq& 5 e_b.
%,\hspace{.5cm} \text{(for }\alpha=e_b\text{)} .
\end{eqnarray}
This linear relation, which can readily be observed in Fig.~\ref{fig-compare1}, is in sharp contrast to the $e_p=e_b$ relation for the BB84 protocol;
specifically, there is a factor of $5$ increase (approximately) in the relation for this three-state protocol.
%We may use the approximate relation in \myeqnref{eqn-ep5eb} to get the key generation rate of $R=1-h(e_b)-h(5e_b)$ and an lower bound of $e_b=0.0425$ which is substantially lower
%than the lower bound of BB84 ($e_b=0.110$ \cite{Shor2000}).
%For the special case $e_b=\alpha$, we can use the approximate relation in \myeqnref{eqn-ep5eb} to get $R=1-h(e_b)-h(5e_b)$.

\section{Key generation rate\label{sec-keyrate}}

In the previous section, we derived two upper bounds on the phase error rate for the three-state protocol;
an optimal one is given by the solution of the problem in \myeqnref{eqn-probsimplified1}, and an approximate one is
given by \myeqnref{eqn-epsol1}.
Using the phase error rate upper bounds, the key generation rate can be readily obtained both for the single-photon source case and for the coherent-source case.
%Although the upper bounds was derived with the assumption of a single-photon source,
Obviously, when comparing the performance of the three-state protocol and the BB84 protocol,
the three-state protocol can only perform as good as, but no better than, the BB84 protocol, since one state is absent in the three-state protocol.
Indeed, as we show in the following, the BB84 protocol is superior to the three-state protocol in the tolerable QBER, the key generation rate, and the maximal secure distance.

\subsection{Single-photon source and perfect detectors}

When a single-photon source and perfect detectors are used,
the key generation rate on the sifted key using
local operations and one-way classical
communications (1-LOCC) can be obtained by applying Shor-Preskill's argument \cite{Shor2000}:
\begin{eqnarray}
\label{eqn-keyrate1}
R &=& 1- H_2(e_b) - H_2(e_p),
\end{eqnarray}
where $e_p$ is either the approximate upper bound in \myeqnref{eqn-epsol1} or the solution of the problem in \myeqnref{eqn-probsimplified1}, and
$H_2(x)=-x\log_2(x)-(1-x)\log_2(1-x)$ is the binary entropy function.
\begin{figure}
\centering
\includegraphics[width=\figlength]{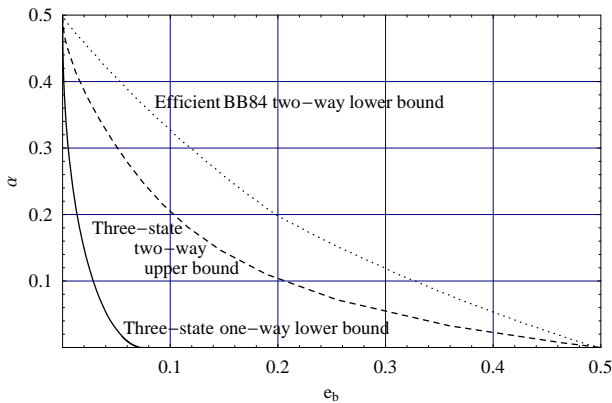}%{secureregion1}
\caption{\label{fig-secureregion1}
%Lower and upper bounds of the protocol.
%The secure region using the approximate upper bound in \myeqnref{eqn-epsol1}.
QBER bounds of the three-state protocol and the efficient BB84 protocol.
For the three-state protocol, the region below the solid curve is secure with 1-LOCC and the region above the dashed curve is insecure with 2-LOCC.
The solid curve is computed using the approximate phase error rate upper bound in \myeqnref{eqn-epsol1},
while the dashed curve is computed using a method based on intercept-and-resend attacks \cite{interceptresend}.
For the three-state protocol, the QBER lower bound for the $Z$-basis states (the $x$-intercept) is $0.075$ and the QBER lower bound for the $\ket{+}$ state (the $y$-intercept) is $1/2$.
%The QBER upper bounds for both
The 2-LOCC lower bound for the efficient BB84 protocol (dotted curve) is reproduced from Ref.~\cite{Ma2006} with a higher precision.
}
\end{figure}
Fig.~\ref{fig-secureregion1} shows the secure region using this key generation rate with the approximate upper bound in \myeqnref{eqn-epsol1}.
The curve is found by searching for $(e_b,\alpha)$ such that the key rate in \myeqnref{eqn-keyrate1} is zero.
The highest tolerable QBER of the data bits is $e_b=0.075$ when $\alpha=0$;
whereas the highest tolerable QBER of the check bits is $\alpha=1/2$ when $e_b=0$.
Also shown in the figure are the upper bound for the protocol with local operations and two-way classical communications (2-LOCC) computed using a method based on intercept-and-resend attacks proposed by us \cite{interceptresend},
and the lower bound for the efficient BB84 protocol \cite{Lo2005b} with 2-LOCC.
The latter is reproduced from Ref.~\cite{Ma2006} (the ``Gottesman-Lo'' curve in Fig.~2 of Ref.~\cite{Ma2006}) with a higher precision.
We can compare the three-state protocol with the efficient BB84 protocol regarding the tolerable bit error rates.
%A lower bound on the tolerable bit error rates (one for the $Z$ basis and another for the $X$ basis) for the efficient BB84 protocol is given by the ``Gottesman-Lo'' curve in Fig.~2 of Ref.~\cite{Ma2006}.
It can be seen that the lower bound curve for the efficient BB84 protocol is above the upper bound curve for the three-state protocol.
% shown in Fig.~\ref{fig-secureregion1} here.
Thus, the efficient BB84 protocol can tolerate higher bit error rates than the three-state protocol.
%The point where $e_b=\alpha$ of the efficient BB84 curve corresponds to the tolerable bit error rate of the BB84 protocol.

We consider the special case $e_b=\alpha$,
which corresponds to a $45$-degree line in Fig.~\ref{fig-secureregion1}.
In this case, we may obtain the tolerable $e_b$ of the three-state protocol from the figure or by
substituting
the approximate relation given in \myeqnref{eqn-ep5eb}
into the key generation rate formula given in \myeqnref{eqn-keyrate1}.
Using the latter method, we obtain
%to get the key generation rate of $R=1-h(e_b)-h(5e_b)$ and
a lower bound of $e_b=0.0425$, which is substantially lower
than the one-way lower bound of BB84 ($e_b=0.110$ \cite{Shor2000}).
The two-way lower bound of BB84
corresponds to the
point where $e_b=\alpha$ on the efficient BB84 curve in Fig.~\ref{fig-secureregion1}, and is equal to $19.9\%$.
This is higher than the two-way upper bound of the three-state protocol at $e_b=\alpha=14.6\%$.
Thus, the BB84 protocol strictly tolerates a higher QBER than the three-state protocol does.
%For the special case $e_b=\alpha$, we can use the approximate relation in \myeqnref{eqn-ep5eb} to get $R=1-h(e_b)-h(5e_b)$.

\subsection{Coherent source and imperfect threshold detectors}

In the previous section,
we  %analyzed the security of
derived the upper bounds on the phase error rate of
the three-state protocol with the assumption of a single-photon source.
Nevertheless, we can easily establish security when a phase-randomized weak coherent light source and imperfect threshold detectors are used by applying the decoy-state method \cite{Hwang2003,Lo2005,Ma2005b,Wang2005a,Wang2005b,Harrington2005,Zhao2006,Zhao2006b,Peng2006,Rosenberg2006}.
In essence, the bit error rates of the single-photon emissions, $e_b$ and $\alpha$, can be upper bounded by the decoy-state method.
The phase error rate of the single-photon emissions, $e_p$, can then be upper bounded
either by using the approximate bound in \myeqnref{eqn-epsol1} or by solving the problem in \myeqnref{eqn-probsimplified1}.
We can further utilize the result of Ref.~\cite{Gottesman2004}, which proves the security of BB84 with an imperfect source, to find the key generation rate of the three-state protocol on the sifted key to be
\begin{equation}
\label{eqn-decoykeyrate}
R \leq -Q_\mu f(E_\mu) H_2(E_\mu) + Q_1 [1-H_2(e_p)],
\end{equation}
where
the subscript $\mu$ denotes the mean photon number for the signal states,
$Q_\mu$ is the gain%
\footnote{The gain of a particular state (e.g. the signal state or the single-photon state) is the probability that
Alice transmits that state and Bob's result is conclusive.}
%the transmitted signal of that type is sent by Alice and results in a conclusive decision by Bob.}
of the signal states, $E_\mu$ is the QBER of the signal states,
$Q_1$ and $e_p$ are the gains and the phase error rates of the single-photon states,
$f(x)$ is the error correction efficiency as a function of error rate, and
$H_2(x)$ is the binary entropy function.
%\footnote{The binary entroy function is defined as $H_2(x)=-x \log_2(x) - (1-x) \log_2(1-x)$.}.
%Here, $E_b$ is the overall bit error rate on all the data qubits, $\Delta$ is the lower bound on the fraction of the single-photon emissions among all emissions detected by Bob, and $e_p$ is the upper bound on the phase error rate of the single-photon emissions.

\begin{figure}
\centering
\includegraphics[width=\figlength]{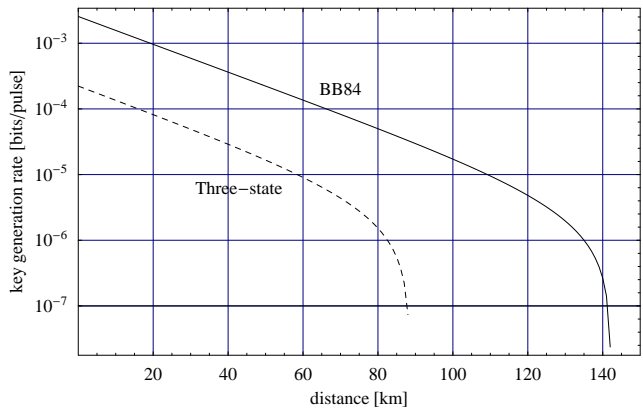}
\caption{\label{fig-decoygraph1}
Comparison between the three-state protocol and the BB84 protocol using the decoy-state method of Ref.~\cite{Lo2005}.
The two key-generation-rate curves are computed from \myeqnref{eqn-decoykeyrate}.
% with $e_p=e_b$ for the BB84 protocol and $e_p=5 e_b$ for the three-state protocol.
The simulation parameters used are from the Gobby-Yuan-Shields (GYS) experiment \cite{Gobby2004} and we have used $f(E_\mu)=1.22$.
The optimal mean photon numbers, $\mu$, for both curves are used at all distances.
The maximal secure distance is $88.5$km for the three-state protocol and $142.2$km for BB84.
}
\end{figure}
Fig.~\ref{fig-decoygraph1} compares the performance of the three-state protocol and the BB84 protocol by using the decoy-state method of Ref.~\cite{Lo2005}.
The simulation parameters used are from the Gobby-Yuan-Shields (GYS) experiment \cite{Gobby2004} and we have used $f(E_\mu)=1.22$.
Here, the phase error rate of the single-photon emissions, $e_p$, is upper bounded by solving the problem in \myeqnref{eqn-probsimplified1}.
As shown in Fig.~\ref{fig-decoygraph1}, the BB84 protocol is better than the three-state protocol in both the key generation rate and the maximal secure distance.
Also, the slopes of both curves can be observed to be approximately the same at short and medium distances.
The difference in the key generation rates for the BB84 protocol and the three-state protocol can be determined from \myeqnref{eqn-decoykeyrate}.
Note that for the BB84 protocol, $e_p=e_b$, while for the three-state protocol, $e_p \approx 5 e_b$ (since we have $e_p=\alpha$ in this model of the QKD setup).
Thus, when the mean photon numbers, $\mu$, for both protocols are the same, the difference in the key generation rates is simply $Q_1 (H_2(5 e_b)-H_2(e_b))$.
On the other hand, when the mean photon numbers are different as in Fig.~\ref{fig-decoygraph1} where the optimal $\mu$ is always used, the difference in the key generation rates has to be calculated directly using \myeqnref{eqn-decoykeyrate}.

\section{Concluding remarks\label{sec-conclusion}}

In this paper, we considered a three-state protocol and proved its security.
Specifically, we showed how the phase error rate of the data bits is upper bounded using the bit error rates observed in the check bits.
This protocol is very similar to the BB84 protocol, sharing the same three qubit states.
Essentially, we showed that, by removing one state from the BB84 protocol and thus destroying the rotational symmetry, the protocol is still secure.
This three-state protocol is interesting in itself since
%it is natural choice for the time-bin-based and
it can be easily implemented in some
frequency-based QKD systems \cite{Molotkov1996,Molotkov1998,Shi2000}.
The result of this work is that these QKD schemes are in fact secure against the most general attacks allowed in quantum mechanics.
%Of course, determining whether these implementations are really secure requires heavy scrutiny on every detail of the implementations.

We compared the three-state protocol with the BB84 protocol both for the single-photon source case and the coherent source case.
Specifically, for the single-photon source case, we showed that the BB84 protocol can strictly tolerate higher bit error rates than the three-state protocol.
For the coherent-source case, the achievable key generation rate and maximal secure distance of the BB84 protocol are both larger than that of the three-state protocol, when the decoy-state method of Ref.~\cite{Lo2005} is used.
In essence, the three-state protocol is inferior to the BB84 protocol;
however, the three-state protocol does have its own merit of being easily implementable in some systems.

We may consider some variations of the three-state protocol.
In the three-state protocol we analyzed, Alice sends states in the $Z$ and $X$ bases with equal probabilities.
This gives rise to Bob using the same basis as Alice with a probability of $1/2$, and thus half of the qubit pairs are discarded.
Although not done in this paper, one may improve on this inefficiency in basis mismatch by applying the idea of efficient BB84 \cite{Lo2005b}.
In the asymptotic limit, Alice and Bob use the same basis with probability approaching one.

In the analysis we provided, we upper bound the phase error rate of the $Z$ basis states by using the average bit error rate of the two $Z$ basis states and the bit error rate of the $\ket{+}$ state.
On the other hand, it is possible to perform a more refined analysis by considering the three bit error rates separately, one for each of the states $\ket{0_z}$, $\ket{1_z}$, and $\ket{+}$.
Although not addressed in this paper, such an analysis can be done in a similar manner as in this paper.
In addition, one may consider a three-state protocol where the check state is not the $\ket{+}$ state but some other state that is an unequal superposition of the $\ket{0_z}$ and $\ket{1_z}$ states,
or a four-state protocol involving the same three states as our three-state protocol plus a state not on the $X$-$Z$ plane of the Bloch sphere (e.g. $(\ket{0_z}+ i \ket{1_z})/\sqrt{2}$).
In this case, it would be interesting to apply the same approach to analyze the security of these protocols.
%a similar analysis can also be performed.

\begin{acknowledgments}
We thank Gilles Brassard, Jim Harrington, Norbert L{\"{u}}tkenhaus, Bing Qi, and Renato Renner for
helpful discussions.
Financial support from
CIAR, NSERC, CIPI, PREA, CRC program, CFI, OIT, OGSST, the Walter Sumner Memorial Fellowship,
and the University of Toronto is
gratefully acknowledged.
HKL thanks Perimeter Institute for a visiting professorship
position where this research is completed.
This research was supported by Perimeter Institute for Theoretical
Physics.
Research at Perimeter Institute is supported in part by the Government
of
Canada through NSERC and by the Province of Ontario through MEDT.
\end{acknowledgments}

\section*{Notes added}

After the first posting of our paper on the arXiv e-print server, we have learnt from Norbert L{\"{u}}tkenhaus
about the existence of an independent proof of
the security of the three-state protocol based on a different approach by
the Geneva group.
Recently, such an independent proof has appeared in
Appendix~A of Ref.~\cite{Branciard2006}.

\appendix

\section{Sufficiency of using one set of $\{a_I,a_X,a_Y,a_Z\}$\label{app-onlyonef}}

In this section, we show that it is sufficient to consider using only one set of $\{a_I,a_X,a_Y,a_Z\}$ in the problem in \myeqnref{eqn-prob1}.
The idea is to construct a new set $\{a_I,a_X,a_Y,a_Z\}$ from two existing sets $\{a_I^{(s)},a_X^{(s)},a_Y^{(s)},a_Z^{(s)}\}$, where $s=1,2$, such that
\begin{eqnarray}
\big|a_\beta\big|^2 &=& \big|a_\beta^{(1)}\big|^2+\big|a_\beta^{(2)}\big|^2, \: \: \beta=I,X,Y,Z  \label{app-onlyonf-eqn1}\\
\big|i a_Y-a_Z\big|^2 &=& \big|i a_Y^{(1)}-a_Z^{(1)}\big|^2+\big|i a_Y^{(2)}-a_Z^{(2)}\big|^2, \label{app-onlyonf-eqn2}\\
\big|a_I+a_X\big|^2 &=& \big|a_I^{(1)}+a_X^{(1)}\big|^2+\big|a_I^{(2)}+a_X^{(2)}\big|^2 \label{app-onlyonf-eqn3}
\end{eqnarray}
are satisfied, meaning that the values of $\{e_b,\alpha,e_p\}$ (c.f. Eqs.~\eqref{eqn-ep1}-\eqref{eqn-alpha1}) are preserved when the new set is used instead of the two existing ones.
Note that condition~\eqref{app-onlyonf-eqn1} already gives the magnitudes of the new $a_\beta$'s.
Thus, only the phases of them remain to be found.
Let us consider the terms with $a_I$ and $a_X$ (the case for $a_Y$ and $a_Z$ are exactly the same).
First note that we can express $|a_I+a_X|^2=|a_I|^2+|a_X|^2+2c|a_I||a_X|$, where $|c|\leq1$ is a function of the phase difference between $a_I$ and $a_X$ and is what we need to determine next.
Once we have found $c$, we can construct the new set by letting $a_X=|a_X|$ and $a_I=\exp(i \arccos c) |a_I|$.

To find $c$, we write condition~\eqref{app-onlyonf-eqn3} as
\begin{eqnarray}
c \big|a_I\big| \big|a_X\big|
=
c^{(1)} \big|a_I^{(1)}\big| \big|a_X^{(1)}\big|+ c^{(2)} \big|a_I^{(2)}\big| \big|a_X^{(2)}\big| ,
\end{eqnarray}
where condition~\eqref{app-onlyonf-eqn1} has been used to eliminate the square terms.
From this, we can readily get
\begin{eqnarray}
c
=
\frac{
c^{(1)} \big|a_I^{(1)}\big| \big|a_X^{(1)}\big|+ c^{(2)} \big|a_I^{(2)}\big| \big|a_X^{(2)}\big|
}
{
%\big|a_I\big| \big|a_X\big|
\sqrt{\big|a_I^{(1)}\big|^2+\big|a_I^{(2)}\big|^2}
\sqrt{\big|a_X^{(1)}\big|^2+\big|a_X^{(2)}\big|^2}
} ,
\end{eqnarray}
where we have again used condition~\eqref{app-onlyonf-eqn1}.
All that is left to do is to verify that $|c|\leq 1$ as follows:
\begin{eqnarray}
|c|
&\leq&
\frac{
\big|a_I^{(1)}\big| \big|a_X^{(1)}\big|+ \big|a_I^{(2)}\big| \big|a_X^{(2)}\big|
}
{
\sqrt{\big|a_I^{(1)}\big|^2+\big|a_I^{(2)}\big|^2}
\sqrt{\big|a_X^{(1)}\big|^2+\big|a_X^{(2)}\big|^2}
}\\
%&=&
%\frac{
%\big|a_I^{(1)}\big| \big|a_X^{(1)}\big|+ g_I\big|a_I^{(1)}\big| g_X\big|a_X^{(1)}\big|
%}
%{
%\sqrt{\big|a_I^{(1)}\big|^2+g_I^2 \big|a_I^{(1)}\big|^2}
%\sqrt{\big|a_X^{(1)}\big|^2+g_X^2 \big|a_X^{(1)}\big|^2}
%}\\
\label{app-onlyonef-eqn-c}
&=&
\frac{1+g_I g_X}
{
\sqrt{1+g_I^2}
\sqrt{1+g_X^2}
} ,
\end{eqnarray}
where the first inequality follows from the fact that $|c^{(1)}|\leq1$ and $|c^{(2)}|\leq1$,
and we have used the definitions $g_I\triangleq\big|a_I^{(2)}\big|/\big|a_I^{(1)}\big|$ and $g_X\triangleq\big|a_X^{(2)}\big|/\big|a_X^{(1)}\big|$.
Now, it is easy to show that the right hand side of \myeqnref{app-onlyonef-eqn-c} is less than or equal to $1$.
For the special case that $\big|a_I^{(1)}\big|=0$ and/or $\big|a_X^{(1)}\big|=0$, the same conclusion of $|c|\leq1$ can be trivially seen.

The construction of the new $a_Y$ and $a_Z$ is similar to the above (which is for $a_I$ and $a_X$), by using conditions \eqref{app-onlyonf-eqn1} and \eqref{app-onlyonf-eqn2}.

\bibliographystyle{apsrev.bst}
\bibliography{paperdb}

\end{document}